\documentclass[twocolumn,showpacs,preprintnumbers,pra,amsmath,amssymb]{revtex4}

\usepackage{graphicx}
\usepackage{bm}
\begin{document}
\title{A Classically Impossible Task Done by Using Quantum Resources}

\author{Won-Young Hwang
}

\affiliation{Department of Physics Education, Chonnam National University, Gwangju 61186, Republic of Korea
}
\begin{abstract}
We propose a task that cannot be done by using any classical mechanical means but can be done with quantum resources. The task is closely related to the violation of Bell's inequality.
\end{abstract}
\maketitle
\section{Introduction}
Quantum resources can perform some tasks that cannot be done by classical ones, e.g., quantum nonlocality \cite{Bel87}, quantum key distribution \cite{Scar09}, and quantum computing \cite{Nie00}.
In particular,  quantum entangled states show marvelous correlations between two remotely separated observers, violating Bell's inequality \cite{Bel87}. In this paper, we propose another classically impossible task that can be done by using quantum mechanics. This is motivated by a recent quantum key distribution protocol with post-selection \cite{Li14} and is
closely related to the violation of Bell's inequality. In the task, two remotely separated persons, Alice and Bob, prepare some (either classical or quantum) physical entities under certain conditions. The physical entities are sent to another person, Charlie, who selects some subsets of the physical entities. Correlations between Alice's and Bob's data corresponding to the selected ones, if the physical entities are classical, can be shown to obey a constraint that is identical to Bell's inequality \cite{Bel87}. However, one can show that the constraint can be violated by using quantum entities.

In Section II, we describe the task in detail. In Section III, we explain why the task cannot be performed by using classical entities while it can be done by using quantum resources. In Section IV, we discuss related issues and present conclusions.
\section{The task}
(i) Alice generates a random number $a$, which we call the basis. Here, $a= 0,1$, and the probability that $a=0$ and $a=1$ are $1/2$ and $1/2$, respectively. She also generates a random number $x$, which we call the state. Here, $x= 0,1$, and the probability that $x=0$ and $x=1$ are $p_{a0}$ and $p_{a1}= 1-p_{a0}$, respectively, for basis $a$. To each combination of the basis $a$ and the state $x$, a physical entity is assigned. Let $(a,x)$ denote the physical entity corresponding to $a$ and $x$. Here, the physical entities may be any physical one, either classical or quantum, and the physical entities may be statistical mixtures of mutually different ones.
Bob does the same things. He generates a random number $b$, which we call the basis. Here $b= 0,1$, and the probability that $b=0$ and $b=1$ are $1/2$ and $1/2$, respectively. Bob also generates a random number $y$, which we call the state. Here, $y= 0,1$, and the probability that $y=0$ and $y=1$ are $p^{\prime}_{b0}$ and $p^{\prime}_{b1}= 1- p^{\prime}_{b0}$,
respectively, for basis $b$. To each combination of the basis $b$ and the state $y$, a physical entity is assigned. Let $(b,y)^{\prime}$ denote the physical entity corresponding to $b$ and $y$.

However, the physical entities must satisfy a condition of basis independence. Suppose that Alice and Bob repeated step (i) many times, generating an ensemble of the physical entities.
Let us hypothetically separate the ensemble of the physical entities according to the basis. Denote the ensemble of Alice's (Bob's) physical entities with basis $a$ ($b$) by $\rho^A_a$ ($\rho^B_b$). For example, $\rho^A_0$ is a statistical mixture of the physical entities $(0,0)$ and $(0,1)$ with corresponding probabilities $p_{00}$ and $p_{01}=1-p_{00}$. {\it The condition is that the statistical mixtures of physical entities corresponding to different bases cannot be discriminated by any physical means; that is, $\rho^A_0$ and $\rho^B_0$ cannot be discriminated from $\rho^A_1$ and $\rho^B_1$, respectively.} Let us take an example of the physical entities satisfying the condition. Suppose that Alice prepares physical entities $(0,0)= |0\rangle$, $(0,1) = |1\rangle$, $(1,0)= |+\rangle$ and $(1,1)= |-\rangle$ with the probabilities $p_{00}= p_{01}= p_{10}= p_{11}= 1/2$. Here, $|0\rangle$ and $|1\rangle$ are two mutually orthogonal states of quantum bits and $|\pm\rangle= (1/\sqrt{2})(|0\rangle \pm |1\rangle)$. Then, the mixtures $\rho^A_0$ and $\rho^A_1$ are given by quantum mixed states described by density operators $(1/2) (|0\rangle \langle 0|+ |1\rangle \langle 1|)$ and $(1/2) (|+\rangle \langle +|+ |-\rangle \langle -|)$, respectively. Because the two density operators are identical, clearly the two mixtures $\rho^A_0$ and $\rho^A_1$  cannot be discriminated by using any physical means; thus, the condition is obeyed.

(ii) Alice and Bob send their respective physical entities $(a,x)$ and $(b,y)$ to Charlie. Then, Charlie may do any physical measurement, including doing nothing, on the physical entities. Based on the measurement's outcomes, Charlie chooses a bit $c$ between $0$ and $1$ and announces it to Alice and Bob.

(iii) Steps (i) and (ii) are repeated many times. Alice and Bob
select only the data for the cases when Charlie announced $c=1$, discarding the others. Let us denote the total number of incidents for $x$ and $y$ with bases $a$ and $b$ by $n(x,y;a,b)$. For example, the total number of incidents when $x=0$ and $y=1$ with bases $a=1$ and $b=0$ is $n(01;10)$. The probabilities for $x$ and $y$ conditioned with bases $a$ and $b$ is given by
\begin{equation}
 p(x,y|a,b)= \frac{n(x,y;a,b)}{\sum_{x,y}n(x,y;a,b)}.
\label{1}
\end{equation}
The correlation between the bases $a$ and $b$ is given by $E(a,b)= p(0,0|a,b)+ p(1,1|a,b)- p(0,1|a,b)- p(1,0|a,b)$. Now, the Bell function $S= E(0,0)+ E(0,1)+ E(1,0)- E(1,1)$ is calculated. If $|S|>2$, then the task is completed.
\section{Tasks that classical entities cannot perform but quantum ones can}
Now, let us see how the task cannot be performed with classical entities. Consider the condition of basis independence that the statistical mixtures of physical entities corresponding to different bases cannot be discriminated by using any physical means; that is, $\rho^A_0$ ($\rho^B_0$) cannot be discriminated from $\rho^A_1$ ($\rho^B_1$). For classical entities, this means that the mixtures $\rho^A_0$ and $\rho^A_1$ are actually identical. Otherwise, the two mixtures can be discriminated because, in principle, classical entities can be directly measured. We can describe classical entities by using a variable $\lambda$, where $0 \leq \lambda \leq 1$, without loss of generality. Mixtures can be characterized by thier probability distributions $P(\lambda)$ with $\int P(\lambda) d\lambda =1 $. Now, the two mixtures $\rho^A_0$ and $\rho^A_1$ must have exactly the same probability distribution $P(\lambda)$.

Let us consider Alice's physical entities $(a,0)$ and $(a,1)$. As said above, they can be mixtures. Let us denote the probability distributions corresponding to $(a,0)$ and $(a,1)$ by $P_{a0}(\lambda)$ and $P_{a1}(\lambda)$, respectively. (Now we have $P(\lambda)= p_{a0} P_{a0}(\lambda)+ p_{a1} P_{a1}(\lambda)$.)
First, we consider the case when $P_{a0}(\lambda)$ and $P_{a1}(\lambda)$ do not overlap. The other case will be dealt with later. Let us consider a $\lambda$ for which $P(\lambda)>0$. Then, we have either $P_{a0}(\lambda)>0$ or $P_{a1}(\lambda)>0$.
Clearly, we can see that in the former (latter) case the physical entity $\lambda$ is from the mixture $(a,0)$ ($(a,1)$). Now, we can classify the set of all $\lambda$'s into four sets:  $\Lambda_{ij}= \{\lambda| P_{0i}(\lambda)>0 \hspace{2mm} \mbox{and} \hspace{2mm} P_{1j}(\lambda)>0  \}$, where $i,j=0,1$. Namely, $\Lambda_{ij}$ is the set of $\lambda$'s that must have come from mixture $(0,i)$ ($(1,j)$) if the basis is $0$ ($1$). For Bob's mixtures $(b,0)^{\prime}$ and $(b,1)^{\prime}$, the same thing can be said. Let us describe Bob's physical entities by using a variable $\lambda^{\prime}$ with $0 \leq \lambda^{\prime} \leq 1$ and characterize the mixtures by using the probability distribution $P^{\prime}(\lambda^{\prime})$ with $\int P^{\prime}(\lambda^{\prime}) d\lambda^{\prime}= 1$. Let us denote the probability distributions corresponding to $(b,0)^{\prime}$ and $(b,1)^{\prime}$ by $P^{\prime}_{b0}(\lambda^{\prime})$ and $P^{\prime}_{b1}(\lambda^{\prime})$, respectively. Here, we also assume that $P^{\prime}_{b0}(\lambda^{\prime})$ and  $P^{\prime}_{b1}(\lambda^{\prime})$ do not overlap.
Now, we also can classify the set of all $\lambda^{\prime}$'s into four sets:  $\Lambda^{\prime}_{kl}= \{\lambda^{\prime}| P^{\prime}_{0k}(\lambda^{\prime})>0 \hspace{2mm} \mbox{and} \hspace{2mm} P^{\prime}_{1l}(\lambda^{\prime})>0  \}$ where $k,l=0,1$.

What happens during the task is that, regardless of basis, Alice and Bob send Charlie certain physical entities $\lambda$ and $\lambda^{\prime}$ chosen according to their probability distributions $P(\lambda)$ and $P^{\prime}(\lambda^{\prime})$, respectively. (See Fig.1)
\begin{figure}
\includegraphics[width=8cm]{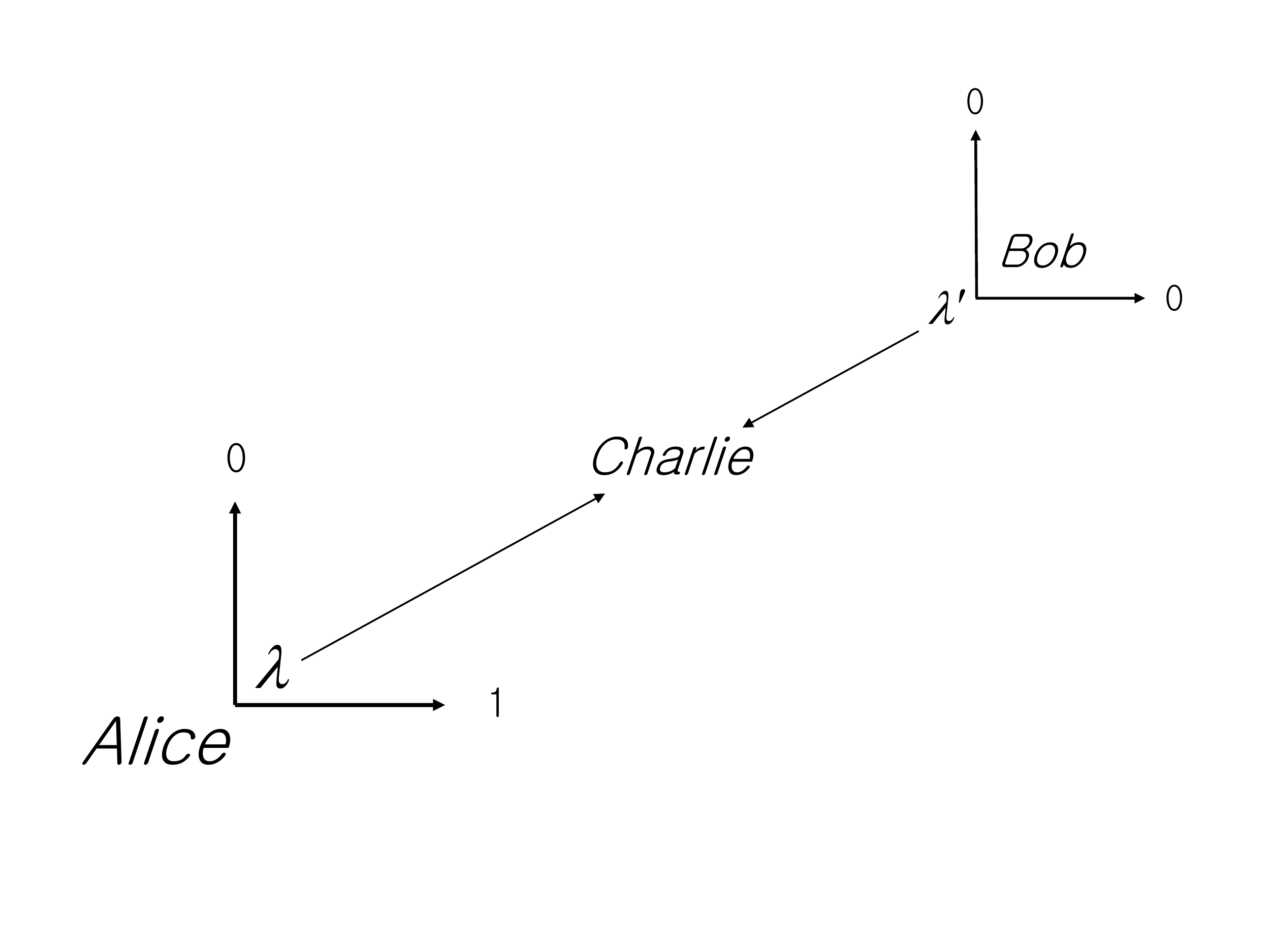}
\caption{Alice and Bob send Charlie certain physical entities $\lambda$ and $\lambda^{\prime}$ chosen according to their probability distributions $P(\lambda)$ and $P^{\prime}(\lambda^{\prime})$, respectively, regardless of the basis. What is considered in this figure is a case when $\lambda \otimes \lambda^{\prime} \in \Lambda_{01}^{00}$, which means that if $\lambda$ came from basis $0$ ($1$), then $\lambda$ is from the $0$ ($1$) state, and that if $\lambda^{\prime}$ came from basis $0$ ($1$), then $\lambda^{\prime}$ is from the $0$ ($0$) state. For a certain $\lambda \otimes \lambda^{\prime}$, Charlie has no information about the basis, so the chosen incidents are statistically evenly distributed over four possible combinations of the basis.}
\label{Fig-1}
\end{figure}
Then, Charlie selects some subset of the product $\lambda \otimes \lambda^{\prime}$ among those he received. Here, we can divide the set of all $\lambda \otimes \lambda^{\prime}$'s into 16 subsets, ${\bf \Lambda}_{ij}^{kl}= \{\lambda \otimes \lambda^{\prime}| \lambda \in \Lambda_{ij} \hspace{2mm} \mbox{and} \hspace{2mm} \lambda^{\prime} \in \Lambda^{\prime}_{kl} \}$. Suppose that steps (i) and (ii) are repeated $N$ times in total, and from among them, $M$  incidents are selected by Charlie in step (iii). Let us denote the number of incidents when $\lambda \in \Lambda_{ij}$ and $\lambda^{\prime} \in \Lambda^{\prime}_{kl}$ among the selected ones by $m(ij;kl)$. The relative frequency is given by
\begin{equation}
 \tilde{m}(ij;kl)= \frac{m(ij;kl)}{\sum_{i,j,k,l}m(ij;kl)}=\frac{m(ij;kl)}{M}.
\label{2}
\end{equation}
Now, let us consider a $\lambda \otimes \lambda^{\prime}$, which  is an element of ${\bf \Lambda}_{ij}^{kl}$ with certain $i,j,k,l$. Note that the $\lambda$ and the $\lambda^{\prime}$ have come from either one of the two bases. $\lambda \otimes \lambda^{\prime} \in {\bf \Lambda}_{ij}^{kl}$ means that if the $\lambda$ is from basis $0$ ($1$), then the $\lambda$ must be from the $i$ ($j$) state, and if the $\lambda^{\prime}$ is from basis $0$ ($1$), then the $\lambda^{\prime}$ must be from the $k$ ($l$) state. However, because of the condition of basis independence, at the step (iii), Charlie has no information about the bases from which the $\lambda$ and the $\lambda^{\prime}$ came. Thus, the probability that a certain $\lambda \otimes \lambda^{\prime}$ is chosen by Charlie is independent of the bases. Combined with the fact that $a,b$ are random, this implies that the $m(ij;kl)$ incidents are statistically evenly distributed over the four combinations of bases; that is, statistically $m(ij;kl)/4$ incidents are the case when the $\lambda \otimes \lambda^{\prime} \in {\bf \Lambda}_{ij}^{kl}$ came from bases $a$ and $b$. Therefore, the relative frequency $\tilde{m}(ij;kl|ab)$ within those incidents when $\lambda \otimes \lambda^{\prime}$ came from the bases $a$ and $b$, is independent of the bases and is equal to the relative frequency $\tilde{m}(ij;kl)$. Now, let us derive the probabilities for $x$ and $y$ conditioned with bases $a$ and $b$, $p(x,y|a,b)$; We can observe that $p(i,k|0,0)= \sum_{j,l} \tilde{m}(ij;kl)$,
$p(i,l|0,1)= \sum_{j,k} \tilde{m}(ij;kl)$, $p(j,k|1,0)= \sum_{i,l} \tilde{m}(ij;kl)$, and $p(j,l|1,1)= \sum_{i,k} \tilde{m}(ij;kl)$. We calculate the correlation $E(0,0)=p(0,0|0,0)+ p(1,1|0,0)- p(0,1|0,0)- p(1,0|0,0)= \sum_{i,j,k,l}\hspace{1mm} (1-2i)(1-2k) \hspace{1mm}\tilde{m}(ij;kl)$, and similarly
$E(0,1)= \sum_{i,j,k,l}\hspace{1mm} (1-2i)(1-2l) \hspace{1mm}\tilde{m}(ij;kl)$,
$E(1,0)= \sum_{i,j,k,l}\hspace{1mm} (1-2j)(1-2k) \hspace{1mm}\tilde{m}(ij;kl)$, and
$E(1,1)= \sum_{i,j,k,l}\hspace{1mm} (1-2j)(1-2l) \hspace{1mm}\tilde{m}(ij;kl)$.
Now, we have the Bell function $S= \sum_{i,j,k,l}\hspace{1mm} \{(1-2i)[(1-2k)+(1-2l)]+(1-2j)[(1-2k)-(1-2l)]\} \hspace{1mm}\tilde{m}(ij;kl)$. Because the absolute value of $\{(1-2i)[(1-2k)+(1-2l)]+(1-2j)[(1-2k)-(1-2l)]\}$ is equal to or less than $2$ in any case and $ \sum_{i,j,k,l} \tilde{m}(ij;kl)=1$, we obtain $|S| \leq 2$. Note that, with respect to the calculations, the derivation here is the same as the one for Bell's inequality.

Now, let us discuss the cases when $P_{a0}(\lambda)$ and $P^{\prime}_{b0}(\lambda^{\prime})$ overlap $P_{a1}(\lambda)$ and  $P^{\prime}_{b1}(\lambda^{\prime})$, respectively. Here, a given physical entity $\lambda \otimes \lambda^{\prime}$ cannot determine which state the physical entities belong to if $\lambda \otimes \lambda^{\prime}$ is in overlapping region. Thus, this case corresponds to the indeterministic local realistic model \cite{Fin82, Sta80}, which still satisfies the Bell inequality. Similarly to the indeterministic case, we can see that the Bell inequality is fulfilled even by the overlapping case,
$S= E(0,0)+ E(0,1)+ E(1,0)- E(1,1)=
\int \{\bar{f}(0,\lambda \otimes \lambda^{\prime})[ \bar{g}(0,\lambda \otimes \lambda^{\prime})+ \bar{g}(1,\lambda \otimes \lambda^{\prime})]
+ \bar{f}(1,\lambda \otimes \lambda^{\prime})[ \bar{g}(0,\lambda \otimes \lambda^{\prime})- \bar{g}(1,\lambda \otimes \lambda^{\prime})]\} P(\lambda) P^{\prime}(\lambda^{\prime}) d (\lambda \otimes \lambda^{\prime}) \leq 2.$
Here, $\bar{f}(a,\lambda \otimes \lambda^{\prime})$ ($ \bar{g}(b,\lambda \otimes \lambda^{\prime})$) is the average of $1-2x$ ($1-2y$) conditioned for basis $a$ ($b$) and $\lambda \otimes \lambda^{\prime}$, respectively. That is, $\bar{f}(a,\lambda \otimes \lambda^{\prime})= \sum_x (1-2x) p(x|a,\lambda \otimes \lambda^{\prime})$ and $\bar{g}(b,\lambda \otimes \lambda^{\prime})= \sum_y (1-2y) p(y|b,\lambda \otimes \lambda^{\prime})$, where, for example, $p(x|a,\lambda \otimes \lambda^{\prime})$ is the probability that the state is $x$ conditioned with that the basis is $a$ and physical entity is $\lambda \otimes \lambda^{\prime}$. At the second equality of this derivation, the condition of basis independence is used.

That quantum resources can perform the task is easy to see. Suppose that Alice prepares the physical entities  $(0,0)= |0\rangle$, $(0,1) = |1\rangle$, $(1,0)= |+\rangle$ and $(1,1)= |-\rangle$ with the probabilities $p_{00}= p_{01}= p_{10}= p_{11}= 1/2$, respectively, and that Bob prepares $(0,0)^{\prime}= |\theta= \pi/4 \rangle$, $(0,1)^{\prime} = |\theta= 5\pi/4\rangle$, $(1,0)^{\prime}= |\theta= 3\pi/4\rangle$ and $(1,1)^{\prime}= |\theta= 7\pi/4\rangle$ with the probabilities $p^{\prime}_{00}= p^{\prime}_{01}= p^{\prime}_{10}= p^{\prime}_{11}= 1/2$, respectively, where $|\theta= \theta \rangle= \cos(\theta/2)|0\rangle+ \sin(\theta/2)|1\rangle$.
This satisfies the condition of basis independence as we have seen in Section II. Charlie performs a measurement composed of $|\varphi^+ \rangle \langle \varphi^+ |$ and $1- |\varphi^+ \rangle \langle \varphi^+|$, where the Bell state is $|\varphi^+ \rangle = (1/\sqrt{2})(|00\rangle + |11\rangle)$, and announces $1$ and $0$ when the outcome is $|\varphi^+ \rangle \langle \varphi^+ |$ and $1- |\varphi^+ \rangle \langle \varphi^+|$, respectively. With the fact that the probability for Charlie to get outcome $|\varphi^+ \rangle \langle \varphi^+ |$ for $|\theta=A\rangle |\theta=A+ \Delta \rangle$ is $(1/2) \cos^2 (\Delta/2)$, one can easily see the $S= 2\sqrt{2} >2$.

\section{Discussion and conclusion.}
If Alice and Bob can perfectly prepare the quantum states as prescribed, then the condition of basis independence is fulfilled because density operators corresponding to different bases are identical. In practice, however, state preparation cannot be perfect although the prepared states may be close to the prescribed ones. Thus, the condition cannot be satisfied perfectly. However, even in practice, a way exists to prepare states satisfying the condition of basis-independence perfectly; Alice  prepares an imperfect state that is close to the Bell state $|\varphi^+ \rangle = (1/\sqrt{2})(|0\rangle_{\alpha} |0 \rangle_{\beta} + |1\rangle_{\alpha} |1\rangle_{\beta})$. Then, she either performs an imperfect measurement that is close the $Z$ measurement composed of $|0\rangle \langle0|$ and $|1\rangle \langle1|$ or performs an imperfect measurement that is close the $X$ measurement composed of $|+\rangle \langle+|$ and $|-\rangle \langle-|$ on the quantum state at the $\alpha$ site. Then, the quantum state prepared at the $\beta$ site, although it is not in the prescribed state perfectly, satisfies the condition of basis independence  perfectly in order to avoid faster-than-light communication, as is well known \cite{Nie00}. Now, similarly Bob prepares his (imperfect) states obeying the condition of basis independence, and Charlie does his (imperfect) measurement that is close to the prescribed one. Then, the $S$ value can still be close to $2\sqrt{2}$ completing the task because the operations are close to the prescribed ones although not perfect. Here, we can see that an experiment on entanglement swapping \cite{Zuk93,Jen02} can be transformed to that on the task proposed here if Alice and Bob perform their measurements earlier than Charlie. (The other case when Charlie performs the measurement earlier corresponds to an experiment that violates Bell's inequality. The two cases are equivalent to each other, as is well known.) Thus, practical implementation of entanglement swapping can be immediately used for the task.

The detection loophole \cite{Pea70} also applies to the task; Let us change the rule of the game by newly introducing a case when the values of the $x$ and $y$ are two. That is, Alice (Bob) chooses one of $0$,$1$, and $2$ for $x$ ($y$) in step (i). Following the same reasonings as before, the set of physical entities $\lambda \otimes \lambda^{\prime}$ are divided into $3^4=81$ subsets ${\bf \Lambda}_{ij}^{kl}$'s. The rule is also changed such that incidents with value $2$ are discarded by Alice and Bob; for example, if $\lambda \otimes \lambda^{\prime} \in {\bf \Lambda}_{02}^{01}$ and Alice's basis $a=1$, then the incidents, even if selected by Charlie, are discarded by Alice and Bob regardless of Bob's basis $b$. Let us take a illustrating example that even achieves $S=4$; the statistical measures of the sets ${\bf \Lambda}_{02}^{02}$, ${\bf \Lambda}_{02}^{20}$, ${\bf \Lambda}_{20}^{02}$, and ${\bf \Lambda}_{20}^{21}$ are all $1/4$ while the statistical measures of other sets are zero.

Let us discuss the potential advantage of our proposal over the Bell test. Because only the incidents selected by Charlie are considered, the task can be performed regardless of Charlie's detector efficiency. Thus, if we can prepare perfect states as prescribed, then we can perform a classically impossible task regardless of the detector's efficiency. If this is the case, then it is an advantage because in the case of a loophole-free violation of Bell inequality, high efficiency detectors are necessary \cite{Ebe93,Hwa96}. Because of the imperfection of the states, however, the condition of basis independence is also imperfectly obeyed. Thus, saying that the task has been performed is difficult. In order to satisfy the condition of basis independence perfectly, we need to adapt the method above, which uses highly entangled states. In this case, however, the detection loophole still exists. Thus, high-efficiency detectors are needed to perform the task.

To summarize, we proposed a task that could not be done by using any classical mechanical means but could be done with quantum resources. Under the condition of basis independence, Alice and Bob prepare some physical entities. They send the entities, while keeping data about the entities, to Charlie, who selects some of the entities. If only classical resources are allowed, the correlation between Alice's and Bob's data corresponding to the selected ones can be shown to satisfy an inequality, that is the same as Bell's inequality. It was shown that quantum resources can violate the inequality. Then, we discussed the issues of imperfect states and the potential advantage of the task.

\section*{Acknowledgement}
This study was supported by Institute for Information and Communications Technology Promotion (IITP) grant funded by the Korea Government (MSIP) (No. R0190-18-2028, Practical and Secure Quantum Key Distribution).



\begin{references}
\bibitem{Bel87} J. S. Bell, {\it Speakable and Unspeakable in Quantum Mechanics} (Cambridge University Press, Cambridge, UK, 1987).
\bibitem{Scar09}
  V. Scarani, B.-P. Helle, N. J. Cerf, M. Dusek, N. L\"utkenhaus, and M. Peev, Rev. Mod. Phys. {\bf 81}, 1301 (2009).

\bibitem{Nie00} M. A. Nielsen and I. L. Chuang, {\it Quantum Computation and Quantum Information} (Cambridge
                University Press, Cambridge, UK, 2000).
\bibitem{Li14} H.-W. Li, Z.-Q. Yin, W. Chen, S. Wang, G.-C. Guo, and Z.-F. Han, Phys. Rev. A {\bf 89}, 032302 (2014).
\bibitem{Fin82} A. Fine, Phys. Rev. Lett. {\bf 48}, 291 (1982).
\bibitem{Sta80} H. P. Stapp, Found. Phys. {\bf 10}, 767 (1980).
\bibitem{Pea70} P. M. Pearle, Phys. Rev. D {\bf 2}, 1418 (1970).
\bibitem{Ebe93} P. H. Eberhard, Phys. Rev. A {\bf 47}, 747 (1993).
\bibitem{Hwa96} W.-Y. Hwang, I. G. Koh, and Y. D. Han, Phys. Lett. A {\bf 212}, 309 (1996).
\bibitem{Zuk93} M. Zukonwski, A. Zeilinger, M. A. Horne, and A. K. Ekert, Phys. Rev. Lett. {\bf 93}, 4287 (1993).
\bibitem{Jen02} T. Jennewein, G. Weihs, J.-W. Pan, and A. Zeilinger, Phys. Rev. Lett. {\bf 88}, 017903 (2002).

\end{references}
\end{document}